\documentclass[a4paper]{jpconf}
\usepackage{graphicx}
\usepackage{upgreek}
\usepackage{bm}
\usepackage{epstopdf}

\begin{document}

\title{Implementation of an offset-dipole magnetic field in a pulsar modelling code}

\author{M Breed$^1$, C Venter$^1$, A K Harding$^2$ and T J Johnson$^{3}$}

\address{$^1$ Centre for Space Research, North-West University, Potchefstroom Campus, Private Bag X6001, Potchefstroom, 2520, South Africa} 
\address{$^2$ Astrophysics Science Division, NASA Goddard Space Flight Center, Greenbelt, MD 20771, USA} 
\address{$^3$ National Research Council Research Associate, National Academy of Sciences, Washington, DC 20001, resident at Naval Research Laboratory, Washington, DC 20375, USA}

\ead{20574266@nwu.ac.za}

\begin{abstract}
The light curves of $\gamma$-ray pulsars detected by the {\it Fermi} Large Area Telescope show great variety in profile shape and position relative to their radio profiles. Such diversity hints at distinct underlying magnetospheric and/or emission geometries for the individual pulsars. We implemented an offset-dipole magnetic field in an existing geometric pulsar modelling code which already includes static and retarded vacuum dipole fields. In our model, this offset is characterised by a parameter $\epsilon$ (with $\epsilon=0$ corresponding to the static dipole case). We constructed sky maps and light curves for several pulsar parameters and magnetic fields, studying the effect of an offset dipole on the resulting light curves. A standard two-pole caustic emission geometry was used. As an application, we compared our model light curves with {\it Fermi} data for the bright Vela pulsar. 
\end{abstract}

\section{Introduction}

The first pulsar was discovered in 1967 by Bell and Hewish \cite{Hewish1968}. Pulsars are identified as compact neutron stars, formed in supernova explosions, that rotate at tremendous rates, and their magnetospheres contain strong electric, magnetic, and gravitational fields \cite{Abdo2010a}. Pulsars emit radiation across the electromagnetic spectrum, including radio, optical, X-ray, and $\gamma$-rays \cite{Becker2007}. We focus on $\gamma$-ray pulsars, specifically the Vela pulsar, which is the brightest persistent GeV source in the $\gamma$-ray sky. The Vela pulsar was detected \cite{Abdo2009a} by the {\it Fermi} Large Area Telescope (LAT) \cite{Atwood2009}, a $\gamma$-ray telescope that was launched in June 2008. {\it Fermi} LAT measures $\gamma$-rays in the energy range between 20 MeV and 300 GeV. The second {\it Fermi} pulsar catalogue \cite{Abdo2013} discussing the properties of 117 $\gamma$-ray pulsars has recently been released. The observed light curves are diverse, and detailed geometric modelling of the radio and $\gamma$-ray light curves may therefore provide constraints on the magnetospheric and emission characteristics. In this paper, we discuss the implementation of an offset-dipole magnetic field in a geometric code, and compare some representative light curves with those observed from Vela.

\section{Model}
\subsection{Emission gap geometry}
\begin{figure}[h!]
	\includegraphics[width=22pc,height=16pc]{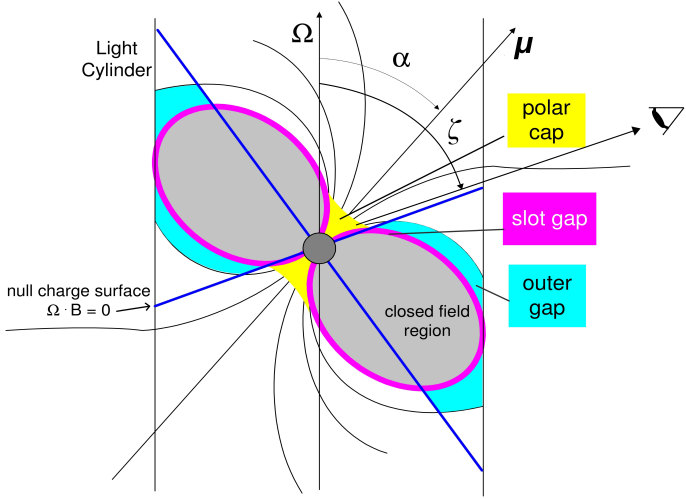}\hspace{0.3cm}
\begin{minipage}[b]{14pc}\caption{\label{geom} A schematic representation of geometric pulsar models. The PC model extends from $R_{\rm{NS}}$ (neutron star radius) up to low altitudes above the surface (yellow region). The TPC emission region (curved magenta lines) extends from $R_{\rm{NS}}$ up to $R_{\rm{LC}}=c/\Omega$ (light cylinder radius), and the OG region (cyan regions) from above the null charge surface (blue lines) to $R_{\rm{LC}}$. Adapted from \cite{Harding2004}. See text for definition of symbols.}
\end{minipage}
\end{figure}

Several models have been used to model $\gamma$-ray emission from pulsars. These include the two-pole caustic (TPC) \cite{Dyks2003} (the slot gap (SG) \cite{Muslimov2003} model may be its physical representation), outer gap (OG) \cite{Cheng1986,Romani1996} and polar cap (PC) model \cite{Daugherty1996}. Consider the ($\bf{\Omega}$, $\boldsymbol{\mu}$) plane, with $\boldsymbol{\mu}$ (the magnetic moment) inclined by an angle $\alpha$ with respect to the rotation axis $\bf{\Omega}$ (the angular velocity). The observer's viewing angle $\zeta$ is the angle between the observer's line of sight and the rotation axis. A `gap region' is defined as the region where particle acceleration and emission take place. The emissivity of $\gamma$-ray photons within this gap region is assumed to be uniform in the corotating frame (for the TPC and OG models) and the $\gamma$-rays are expected to be emitted tangentially to the local magnetic field in this frame \cite{Dyks2004}, which means that the assumed magnetic field geometry is very important with respect to the predicted light curves. The gap region for the TPC model extends from the surface of the neutron star along the entire length of the last closed magnetic field lines, up to the light cylinder (where the corotation speed equals the speed of light), as indicated by the magenta region in Figure~\ref{geom}. For the OG model, the gap region extends from the null-charge surface, where the Goldreich-Julian charge density $\rho_{\rm GJ}=0$ \cite{GJ1969}, to the light cylinder, as indicated by the cyan region. The PC gap (yellow region) extends from the neutron star surface to the low-altitude pair formation front, where the $E$-field is screened by pairs formed via single-photon pair production. In what follows, we will focus on the TPC model.  

\subsection{Magnetic field structure}\label{offset-dipole}
Several magnetospheric structures have been studied, including the static dipole field \cite{Griffiths1995}, the retarded dipole field \cite{Deutsch1955} (a rotating vacuum magnetosphere which can in principle accelerate particles but do not contain any charges or currents) and the force-free field \cite{Contopoulos1999} (being filled with charges and currents, but unable to accelerate particles, since the $E$-field is screened everywhere). A more realistic pulsar magnetosphere \cite{Kalapotharakos2012} would be one that is intermediate between the vacuum retarded and the force-free fields. 

The main focus of this paper is on the offset-dipole $B$-field. Retardation of the $B$-field and asymmetric currents may cause small distortions in the $B$-field structure, shifting the polar caps by small amounts and different directions. In the `symmetric case', in the corotating magnetic frame (where ${\bf{\hat{z}^\prime}}\parallel{\boldsymbol{\mu}}$), the offset-dipole $B$-field in spherical coordinates is given by \cite{Harding2011}
\begin{equation}
\mathbf{B}^\prime(r^\prime,\theta^\prime,\phi^\prime) = \frac{\mu}{{\it r}^{\prime3}}\left[\cos\theta^\prime\mathbf{\hat{r}^\prime} + \frac{1}{2}(1+a)\sin\theta^\prime\bm{\hat{\theta}^\prime} - \epsilon\sin\theta^\prime\cos\theta^\prime\sin(\phi^\prime-\phi_0)\bm{\hat{\phi}^\prime}\right],
\end{equation}
where $\mu=B_{0}R_{\rm NS}/2$ is the magnetic moment, $B_0$ the surface $B$-field strength at the magnetic pole, $R_{\rm NS}$ the stellar radius, $\phi_0$ the magnetic azimuthal angle defining the plane in which the offset occurs, and 
\begin{equation}
a = \epsilon\cos(\phi^\prime-\phi_0).
\end{equation}
The magnitude of the offset is characterised by a parameter $\epsilon$, which represents a shift of the polar cap from the magnetic axis, with $\epsilon=0$ corresponding to the static-dipole case.  

\section{Implementation of the offset-dipole $B$-field in the code}
\subsection{Transformation of $B$-field}
\begin{figure}[t]
\begin{center}
\includegraphics[width=14cm]{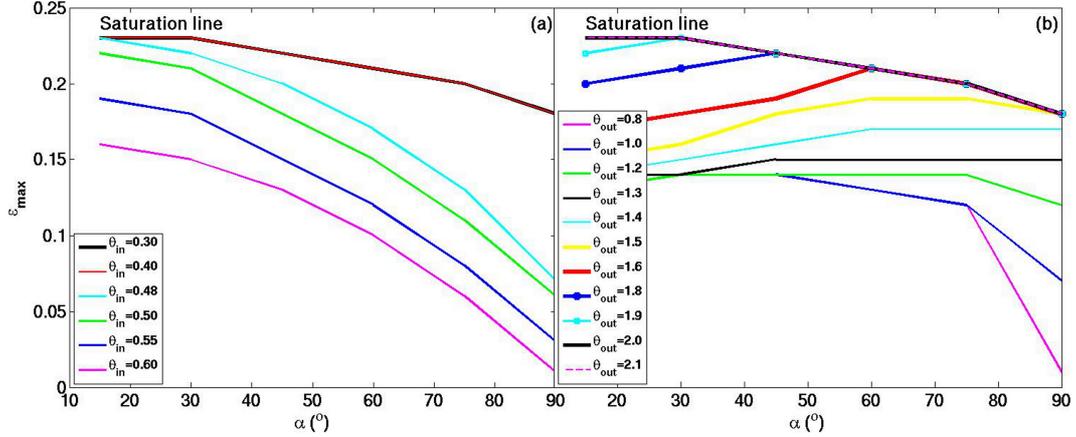}
\end{center}
\caption{\label{Thout} A plot of $\epsilon_{\rm max}$ as a function of $\alpha$. In panel (a) each line corresponds to a different value of $\theta_{\rm in}$ for a constant value of $\theta_{\rm out}=2.0$ (both in units of polar cap angle $\theta_{\rm PC}\approx\sqrt{R_{\rm NS}/R_{\rm LC}}$). For values smaller than $\theta_{\rm in}=0.40$, $\epsilon_{\rm max}$ becomes saturated, and for larger $\theta_{\rm in}$, $\epsilon_{\rm max}$ decreases significantly. In panel (b) each line corresponds to a different value of $\theta_{\rm out}$ for a constant value of $\theta_{\rm in}=0.3$, where $\epsilon_{\rm max}$ becomes saturated at $\theta_{\rm out}=2.0$, and as $\theta_{\rm out}$ decreases, $\epsilon_{\rm max}$ also decreases.}
\end{figure} 
We start with a $B$-field defined in the magnetic frame (${\bf{\hat{z}^\prime}}\parallel{\boldsymbol{\mu}}$), specified using spherical coordinates 
\begin{equation}\label{eq:rtpprime}
{\mathbf B}^\prime(r^\prime,\theta^\prime,\phi^\prime)=B_r^\prime(r^\prime,\theta^\prime,\phi^\prime)\mathbf{\hat{r}^\prime}+B_\theta^\prime(r^\prime,\theta^\prime,\phi^\prime)\bm{\hat{\theta}^\prime}+B_\phi^\prime(r^\prime,\theta^\prime,\phi^\prime)\bm{\hat{\phi}^\prime}.
\end{equation}
(We choose $\phi^\prime=0$ in the direction toward $\bm{\Omega}$). We transform this to a Cartesian coordinate system: 
\begin{equation}\label{eq:xyzprime}
\mathbf{B}^\prime(x^\prime,y^\prime,z^\prime)=B_x^\prime(x^\prime,y^\prime,z^\prime)\mathbf{\hat{x}^\prime}+B_y^\prime(x^\prime,y^\prime,z^\prime)\mathbf{\hat{y}^\prime}+
B_z^\prime(x^\prime,y^\prime,z^\prime)\mathbf{\hat{z}^\prime}.
\end{equation}
This is done using expressions that specify spherical unit vectors and coordinates in terms of Cartesian coordinates (see, e.g., \cite{Griffiths1995}). Next, we rotate both the $B$-field components and the associated Cartesian frame (or equivalently, the position vector) through an angle $-\alpha$, thereby transforming the $B$-field from the magnetic frame to the rotational frame (${\bf{\hat{z}}}\parallel{\boldsymbol{\Omega}}$): 
\begin{equation}\label{eq:xyz}
\mathbf{B}(x,y,z)=B_x(x,y,z)\mathbf{\hat{x}}+B_y(x,y,z)\mathbf{\hat{y}}+B_z(x,y,z)\mathbf{\hat{z}}.
\end{equation}
After initial implementation of the offset-dipole field in the geometric code we discovered that we could solve the polar cap rim (for details, see \cite{Dyks2004}) only for small values of $\epsilon$. We improved the range of $\epsilon$ by changing the parameters $\theta_{\rm in}$ and $\theta_{\rm out}$, which delimit a bracket in colatitude thought to contain the last open field line (tangent to $R_{\rm LC}$). Figure~\ref{Thout} indicates the progressively larger range of $\epsilon$ that we were able to use upon decreasing  $\theta_{\rm in}$ and increasing $\theta_{\rm out}$.  

\subsection{The offset-dipole E-field}  
\begin{figure}[b]\label{RR}
\begin{center}
\includegraphics[width=14cm]{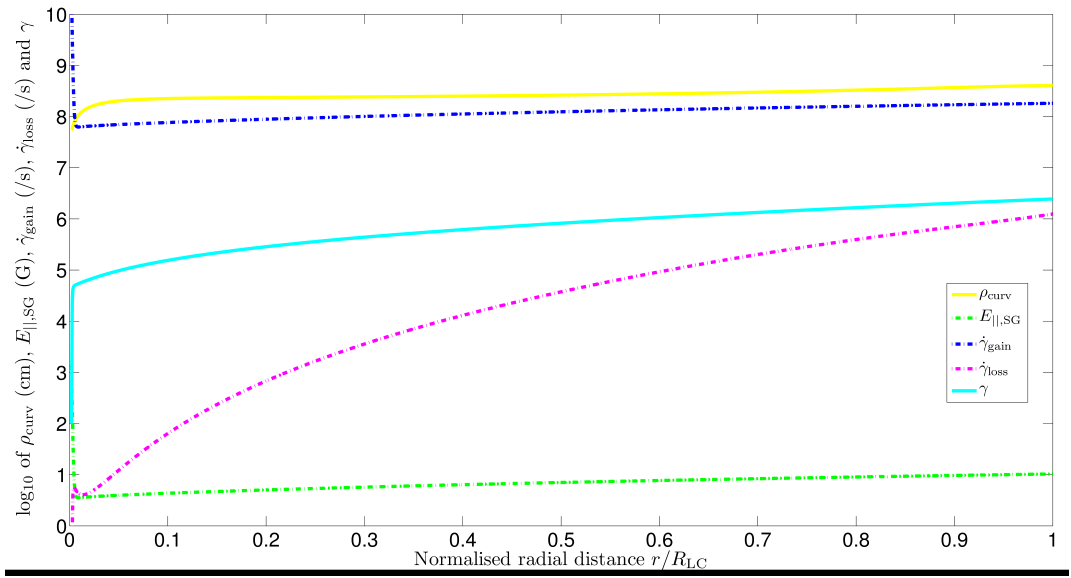}
\end{center}
\caption{\label{RR} Plot of log$_{10}$ of curvature radius $\rho_{\rm curv}$ (solid yellow line), general $E_{\parallel,{\rm SG}}$-field (dash-dotted green line), gain (acceleration) rate $\dot{\gamma}_{\rm gain}$ (dash-dotted dark blue line), loss rate $\dot{\gamma}_{\rm loss}$ (dash-dotted pink line), and the Lorentz factor $\gamma$ (solid light blue line) as a function of normalised radial distance $r/R_{\rm LC}$. We used $P=0.0893$ s, $B_0=1.05\times{10}^{13}$ G (corrected for general relativistic effects), $I=0.4MR_{\rm NS}^2=1.14\times10^{45}$ g\,cm$^2$, $\phi^{\prime}=1.60$ radians (at the stellar surface), $\xi_{\ast}= 0$, and $\eta_c=1.4$.}
\end{figure} 
It is important to take the accelerating $E$-field into account (in a physical model) when such expressions are available, since this will modulate the emissivity in the gap (as opposed to geometric models where we just assume constant emissivity in the corotating frame). The low-altitude $E$-field in the offset-dipole magnetosphere, for the SG model, is given by
\begin{equation}\label{eq:E_low}
E_{\parallel,{\rm low}}\approx{-3}{\mathcal{E}_0}\nu_{\rm SG}x^a\left\{\frac{\kappa}{\eta^4}e_{\rm 1A}\cos\alpha+\frac{1}{4}\frac{\theta_0^{1+a}}{\eta}\Bigg[e_{\rm 2A}\cos\phi^{\prime}+\frac{1}{4}\epsilon\kappa{e_{\rm 3A}}(2\cos\phi_0-\cos(2\phi^{\prime}-\phi_0))\Bigg]\sin\alpha\right\}(1-\xi_\ast^2), 
\end{equation}
where the symbols have the same meaning as in \cite{Muslimov2004}. Here 
\begin{equation}\label{eq:e1A}
e_{\rm 1A}=1+\frac{a}{3}(\eta^3-1);\quad e_{\rm 2A}=(1+3a)\eta^{(1+a)/2}-2a;\quad e_{\rm 3A}=\frac{5-3a}{\eta^{(5-a)/2}}+2a.
\end{equation}
We approximate the high-altitude SG $E$-field by \cite{Muslimov2004}
\begin{equation}\label{eq:E_high}
E_{\parallel,{\rm high}}\approx-\frac{3}{8}\Bigg(\frac{\Omega{R}}{c}\Bigg)^3\frac{B_{\rm 0}}{f(1)}\nu_{\rm SG}x^a\left\{\Bigg[1+\frac{1}{3}\kappa\Bigg(5-\frac{8}{\eta^3_c}\Bigg)+2\frac{\eta}{\eta_{\rm LC}}\Bigg]\cos\alpha+\frac{3}{2}\theta_{0}H(1)\sin\alpha\cos\phi^{\prime}\right\}(1-\xi_\ast^2),
\end{equation}
and the general $E$-field valid from $R_{\rm NS}$ to $R_{\rm LC}$ by
\begin{equation}\label{eq:E_total}
E_{\parallel,{\rm SG}}{\simeq}E_{\parallel,{\rm low}}\exp\Bigg(\frac{-(\eta-1)}{(\eta_{c}-1)}\Bigg)+E_{\parallel,{\rm high}},
\end{equation}
where $\eta_c=r_c/R_{\rm NS}$ is the critical scaled radius where the high-altitude and low-altitude $E$-field solutions are matched (see, e.g., equation [59] of \cite{Muslimov2004}).
\begin{figure}[b]
\begin{center}
\includegraphics[width=14cm]{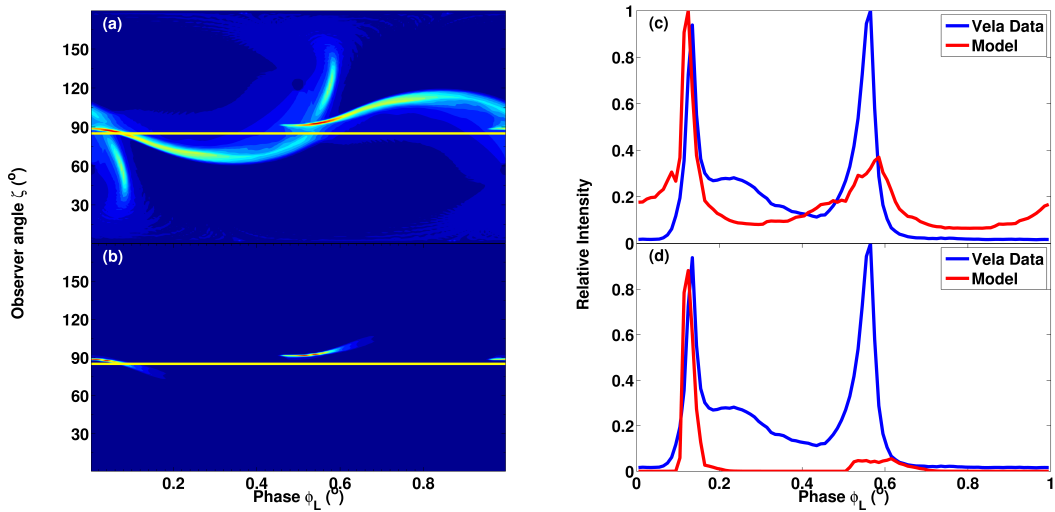}
\end{center}
\caption{\label{const} Phaseplots and light curves for the offset-dipole field. Panels (a) and (b) illustrate the emission per solid angle versus $\zeta$ and $\phi_{\rm L}$ for $\alpha=60^\circ$ and panels (c) and (d) their corresponding light curves for $\zeta=85^\circ$ (indicated by the solid yellow lines). The solid blue line denotes the observed Vela profile~(e.g., \cite{Abdo2013}) and the solid red line our model profile. Panel (a) and (c) are for the case of constant emissivity, whereas panel (b) and (d) are for the case when using the offset-dipole $E$-field and solving for $\gamma$. We assumed $\epsilon=0.2$, a gap width of 5\% of the polar cap angle, and $R_{\rm max}=1.2R_{\rm LC}$ for an SG model.}
\end{figure}

\section{Results}
\subsection{Solution of particle equation of motion}
Using equation~(\ref{eq:E_total}) we next solve the particle transport equation (taking only curvature radiation losses into account)
\begin{equation}
\dot{\gamma}=\dot{\gamma}_{\rm gain}+\dot{\gamma}_{\rm loss}=\frac{eE_{\parallel,{\rm SG}}}{mc}-\frac{2e^2\gamma^4}{3\rho^2_{\rm curv}mc},
\end{equation}
to obtain the particle Lorentz factor $\gamma(\eta,\phi^{\prime},\xi_{\ast})$ with $\eta=r/R_{\rm LC}$ the normalised radial distance, and $\xi_{\ast}$ a normalised colatitudinal angle which is $\xi_{\ast}=0$ at the middle of the SG and $\xi_{\ast}=1$ at the boundaries \cite{Muslimov2003}. Radiation reaction occurs when the energy gain balances the losses, and $\dot{\gamma}=0$. In Figure~\ref{RR} we plot the log$_{10}$ of curvature radius $\rho_{\rm curv}$, general $E$-field $E_{\parallel,\rm SG}$, gain rate $\dot{\gamma}_{\rm gain}$, loss rate $\dot{\gamma}_{\rm loss}$, and particle Lorentz factor $\gamma$ as a function of $\eta$. We can see that the radiation reaction limit is not reached in this case, due to the relatively low SG $E$-field. The Lorentz factor is initially set to $\gamma=100$ and rapidly rises until it nearly reaches $\gamma\sim{10^6}$. For different choices of $\phi^{\prime}$ and $\xi_{\ast}$, the $E$-field may be even lower and $\gamma$ may not even exceed $\sim10^5$, leading to negligible curvature radiation losses along those field lines. On `unfavourably curved' field lines ($\phi^\prime\approx\pi$), the $E$-field may even change sign at higher altitudes. This may cause oscillation of particles and very low values of $\gamma$, and such field lines should be ignored when constructing phaseplots. 

\subsection{Phaseplots and light curves}
In Figure~\ref{const}, we show the phaseplots (emission per solid angle versus $\zeta$ and observer phase $\phi_{\rm L}$) and the corresponding light curves (i.e., cuts along constant $\zeta$) for the offset-dipole $B$-field and TPC model. The dark circle in panel (a) is the non-emitting polar cap, and the sharp, bright regions are the emission caustics, where radiation is bunched in phase due to relativistic effects. The caustic structure is qualitatively different between the two cases (constant emissivity vs.\ solution of $\gamma$ using $E_\parallel$), leading to differences in the resulting light curves. The caustics seem wider and more pronounced in the constant-emissivity case. The blue lines in panel (c) and (d) are the measured Vela profiles \cite{Abdo2013}, while the red lines are the predicted light curves. Note that the latter are merely representative, and still fail to adequately reproduce the second large peak of the measured profile.

\section{Conclusions and future work}
We have studied the effect of implementing the offset-dipole $B$-field on $\gamma$-ray light curves for the TPC geometry. We observe that the polar cap is indeed offset compared to the case of the static dipole (not shown Figure~\ref{const}) when assuming a constant emissivity. However, when including an $E$-field and solving for $\gamma$, we see that the resulting phaseplot becomes qualitatively different, given the fact that $\gamma$ only becomes large enough to yield significant curvature radiation at large altitudes. Furthermore, we do not attain the radiation-reaction limit, due to a relatively low $E$-field.
In future, we want to solve for $\eta_c$ on each field line, instead of using a constant value where we match $E$-field solutions. Lastly, we want to produce light curves for several model parameters and search for a best-fit profile, thereby constraining Vela's low-altitude magnetic structure and system geometry.

\ack
This work is supported by the South African National Research Foundation (NRF). AKH acknowledges the support from the NASA Astrophysics Theory Program. CV, TJJ, and AKH acknowledge support from the {\it Fermi} Guest Investigator Program.


\section*{References}
\begin{thebibliography}{9}
\bibitem{Hewish1968} Hewish A {\it et al.} 1968 {\it Nature} {\bf 217} 709--13
\bibitem{Abdo2010a} Abdo A A {\it et al.} 2010 {\it ApJS} {\bf 187} 460--94
\bibitem{Becker2007} Becker W, Gil J A and Rudak B 2007 {\it Highlights of Astronomy} {\bf 14} 109--38
\bibitem{Abdo2009a} Abdo A A {\it et al.} 2009 {\it ApJ} {\bf 696} 1084--93
\bibitem{Atwood2009} Atwood W B {\it et al.} 2009 {\it ApJ} {\bf 697} 1071--102
\bibitem{Abdo2013} Abdo A A {\it et al.} 2013 {\it ApJS} {\bf 208} 17--76
\bibitem{Dyks2003} Dyks J and Rudak B 2003 {\it ApJ} {\bf 598} 1201--6
\bibitem{Muslimov2003} Muslimov A G and Harding A K 2003 {\it ApJ} {\bf 588} 430--40
\bibitem{Cheng1986} Cheng K S, Ho C and Ruderman M 1986, {\it ApJ} {\bf 300} 500--39
\bibitem{Romani1996} Romani R W 1996 {\it ApJ} {\bf 470} 469--78
\bibitem{Daugherty1996} Daugherty J K and Harding A K 1996 {\it ApJ} {\bf 458} 278--92 
\bibitem{Harding2004} Harding A K 2004 {\it 22nd Texas Symp. on Relativistic Astrophysics} ed P Chen, E Bloom {\it et al.} p 40
\bibitem{Dyks2004} Dyks J, Harding A K and Rudak B 2004 {\it ApJ} {\bf 606} 1125--42
\bibitem{GJ1969} Goldreich P and Julian W H 1969 {\it ApJ} {\bf 157} 869--80
\bibitem{Griffiths1995} Griffiths D J 1995 {\it Introduction to Electrodynamics} (San Francisco: Pearson Benjamin Cummings)
\bibitem{Deutsch1955} Deutsch A J 1955 {\it Annales d'Astrophysique} {\bf 18} 1--10
\bibitem{Contopoulos1999} Contopoulos I, Kazanas D and Fendt C 1999 {\it ApJ} {\bf 511} 351--8
\bibitem{Kalapotharakos2012} Kalapotharakos C, Kazanas D, Harding A and Contopoulos I 2012 {\it ApJ} {\bf 749} 1--15
\bibitem{Harding2011} Harding A K and Muslimov A G 2011 {\it ApJ} {\bf 743} 181--96
\bibitem{Muslimov2004} Muslimov A G and Harding A K 2004 {\it ApJ} {\bf 606} 1143--53
\end{thebibliography}
\end{document}